# Explainable Software Defect Prediction from Cross Company Project Metrics Using Machine Learning

Susmita Haldar
*School of Information Technology*
*Fanshawe College*
London, Canada
shaldar@fanshawec.ca

Luiz Fernando Capretz
*Department of Electrical and Computer Engineering*
*Western University*
London, Canada
lcapretz@uwo.ca

*Abstract*—Predicting the number of defects in a project is critical for project test managers to allocate budget, resources, and schedule for testing, support and maintenance efforts. Software Defect Prediction models predict the number of defects in given projects after training the model with historical defect related information. The majority of defect prediction studies focused on predicting defect-prone modules from methods, and class-level static information, whereas this study predicts defects from project-level information based on a cross-company project dataset. This study utilizes software sizing metrics, effort metrics, and defect density information, and focuses on developing defect prediction models that apply various machine learning algorithms. One notable issue in existing defect prediction studies is the lack of transparency in the developed models. Consequently, the explain-ability of the developed model has been demonstrated using the state-of-the-art post-hoc model-agnostic method called Shapley Additive exPlanations (SHAP). Finally, important features for predicting defects from cross-company project information were identified.

## I. Introduction

As the complexity of software increases, delivering quality and defect-free software seems challenging due to strict timeline, and limited budget. Managers strives to identifies the defects as early as possible in the software development life cycle because addressing defects in later stages can incur higher costs [1]. However, Identifying the potential number of defects in a project takes significant effort. This brings the need to have an automated process for predicting bugs. Defect prediction using machine learning is an emerging technology that leverages human experience by automating manual efforts to anticipate defects in software systems [2], [3], [4]. The majority of the SDP studies focused on identifying defect-prone vs. non-defect-prone modules. However, it is equally important to consider the number of defects as fixing a single defect may be less costly and require less resources compared to fixing a large quantity of defects.

Obtaining real-world data for training machine learning models for defect prediction has limitations. Companies may be hesitant to share proprietary information regarding the actual number of defects found after project delivery. In addition, project managers may lack the technical skill-set to comprehend the predicted outcome. By understanding the reasoning behind a prediction, project managers could select and adjust the selected attributes for assessing the need to allocate more resources in these identified projects with high number of defects.

Predicting defects from cross-company project information is important as conducting study on the same type of projects may not yield a reliable model due to differences in project types or data collection sources in real-world problems. Additionally, historical information may not be available for the same type of projects. Bai et al. [5] investigated the issues of transfer learning in cross-project defect prediction and proposed a three-stage weighted framework for multi-source transfer learning process-based SDP model. It is also important to understand which features to be extracted for building an effective SDP model. Recently, Balogun et al. [6] addressed the fact that the high dimensionality of software metric features can affect the performance of SDP models. They conducted feasibility studies on feature selection of reliable SDP model by applying hybrid feature selection algorithms.

This study will contribute to software engineering research domain by answering to the following research questions:

RQ1: Can we build a SDP model from generic cross-company project-level information without segregating the project level information based on project size or development types? RQ2: Are defect density and software size strong predictors for predicting the number of software defects in a project? RQ3: Can we identify at least three features from this cross-company project dataset that are important for predicting defects based on similar metrics? RQ4: Can we interpret the predicted number of defects from the developed SDP model?

This paper is organized into several sections. The related work on defect prediction using machine learning is presented in section II. This is followed by the methodology used in this paper in section III. The SDP models developed using ML algorithms, and the results are presented in section IV. Section V summarizes the result analysis and discussion. Threats to the validity of our work are presented in Section VI. Finally, the conclusion and future work are described in section VII.

## II. BACKGROUND AND RELATED WORK

Software defect prediction(SDP) has emerged as a popular research topic over the last several decades [3], [6], [7]. Researchers have utilized various classification techniques to build these models including Logistic Regression [8], Naïve Bayes classifier [9], Support Vector Machine [8], Artificial Neural Networks [10], Decision Tree Classifiers [11], Random Forest Algorithms [12], kernel PCA [13], Deep Learning [14], combination of Kernel PCA and Deep Learning [15] [16] and ensemble learning techniques [17] etc. Aleem et al. [3] explored different machine learning techniques for software bug detection and provided a comparative performance analysis of these algorithms.

Several studies used discretizing continuous defect counts into defective and non-defective modules for SDP models [18], [19]. However, binning the continuous data as independent variable may lead to information loss that can affect the performance and interpretation of SDP models [20]. Rajbahadur et al. [20] recommended that future SDP studies should consider building regression-based classifiers. In this study, we have used regression-based machine learning techniques for predicting the total number of defects.

Felix and Lee [21] proposed certain SDP models constructed using code design complexity, defect density, defect introduction time and defect velocity. Their results indicate that the number of defects shows a strong positive correlation with the average defect velocity, but a weak positive correlation with the average defect density and a negative correlation with the average defect introduction time. However, in this work, we can observe a significantly positive relationship with defect density, and number of defects.

In recent years, the need for explainability in machine learning models has gained prominent importance. Gezici and Tarhan [22] utilized three existing model-agnostic-based techniques referred to as EL5, SHAP and LIME to develop an explainable defect prediction model based on gradient boost algorithm classifier. We will explain our SDP model with SHAP on various classifiers as the cost computation for this dataset is reasonable.

In 2017, Almakadmeh et al. [23] analyzed the ISBSG MS-Excel based dataset on Six Sigma measurements and found that the ISBSG dataset has a high ratio of missing data within the data fields of the "Total Number of Defects" variable. They identified that this missing ratio represents a serious challenge when the ISBSG dataset is being used for software defect estimation. To overcome this challenge, along with other cleaning criteria, we have removed the records with missing values in the "Total Number of Defects variable".

Fadi and Al-Manai [24] found a weak correlation between size and defects when conducting a study on these variables. However, our SDP model contradicts this study as functional size shows a significant correlation with the total number of defects in a project. Researchers can focus on collecting size-based metrics from various projects to assist project managers in determining the estimation of the number of defects for scheduling and allocating testing resources.

He et al. [25] collected data from several open-source projects which provided them with information about faulty vs. non-faulty modules for cross-project defect prediction. Unlike their study, we are focusing on predicting the total number of defects instead of just identifying defective vs. non-defective modules. This approach will provide project managers with an approximate number of defects.

Shao et al. [26] conducted research on cross-company project data for building SDP model for ensuring software security and reliability. Their study was facing limitation of conducting research with only part of NASA and PROMISE datasets, and they highlighted the need for collecting more datasets to verify the effectiveness.

Shin et al. [27] showed the existing explain-ability studies using model-agnostic techniques exhibits inconsistencies in explaining the SDP models. Our contribution in this paper includes verifying if we can see alignment in the major contributing features.

## III. METHODOLOGY

### A. Dataset

In this paper, the ISBSG Developments & Enhancements 2021 Release 2 [28] dataset was explored. This original data repository contains a total of 10600 records with 254 features encompassing a broad range of projects across various industry sectors and business areas. Also, this dataset contains projects from over thirty countries [29] worldwide. These features have different groupings based on application types, organization sectors, development types, development environments, scheduling, programming languages, documentation, tools, and methodologies used in the projects etc. as part of various project metrics, size metrics, effort metrics, defect density, quality metrics, effort and other relevant metrics [30], [31].

### B. Feature Extraction and pre-processing

This dataset contains many missing values, and not all fields are required for our study. We have applied filtering to retain features that have an acceptable number of records without being highly correlated. A snapshot of the data filtering technique has been shown in Table I. This feature extraction process involved multiple steps. **Step 1** involved removing records with a blank value in total number of defects field as this feature serves our target variable, and after this step a total number of 2103 records remained. Next, in **step 2**, a field called "age" was derived by subtracting the project implementation date from the current date to assess if the maturity of the project can contribute to the development of SDP model. In **step 3**, to select records with high quality data for building a trustworthy SDP model, we removed records with poor ratings. According to ISBSG

| Field name | Description | Data type |
|---|---|---|
| Industry Sector | Various organization types such as Finance, Communications, and Government etc. | Categorical values |
| Development_Type | Explain if the project was a new development, enhancement or re-development. | Categorical values |
| Primary Programming Language | The primary language used for the development: JAVA, C++, PL/1, Natural, Cobol etc. | Categorical values |
| Count Approach | Techniques used to size the selected projects. For most projects in the ISBSG repository this was the Functional Size Measurement Method (FSM Method) used to measure the functional size such as IFPUG, MARK II, NESMA, FiSMA, COSMIC-FFP etc. | Categorical values |
| Functional Size | The unadjusted function point count. | 64 bit floating-point number |
| Relative Size | Categories of functional size such as large, medium2, medium 1, small, extra small,, extra large etc. categories by 'L', 'M2', 'M1', 'S', 'XS', 'XXS', 'XL' | Categorical values |
| Normalised Work Effort | Full life-cycle effort for all teams reported. | 64 bit floating-point number |
| Defect Density | Number of defects delivered in unit size of software. It is defined as the number of defects per 1000 Functional Size Units of delivered software, in the first month of use. | 64 bit floating-point number |
| 1st Language | This is the primary technology programming language used to build or enhance the software. | Categorical values |
| Age | A derived value from project implementation date by subtracting the project implementation date from current date. | 64 bit floating-point number |
| Total Defects Delivered | Total number of defects reported in the first month of use of the software. | 64 bit floating point number |

Fig. 1. Final dataset with description.

, a quality rating of C or D indicates that the integrity of the data could not be assessed or achieved little credibility. After this cleaning step, a total of 1542 records remained. **Step 4** dealt with removing irrelevant information such as project ID, rating related fields etc. **Step 5** involved finding records with more than 10% of missing values in the column values. Except for the "programming language" field, all other columns with missing values ranging from 1 to 10% were taken out. The missing values in the programming language field were filled with the value of 'unknown'. After this cleaning step, we were left with 12 columns. **Step 6** involved removing records with highly internally correlated values where the correlation exceeded 70%. Fig. 2 shows the correlation matrix among the remaining non-categorical predictors in this dataset. From this figure, we can observe that summarized work effort and adjusted functional points are highly correlated with functional size and normalized effort. Consequently, these two features were removed.

The final set of features selected from this dataset is shown in Fig. 1. The resulting dataset consists of 1254

TABLE I
PRIMARY CRITERIA USED FOR SELECTING RECORDS

| Criteria | Field description | Actions taken |
|---|---|---|
| Data Quality Rating | This field contains an ISBSG rating code of A, B, C or D applied to the project data | Filtered record based on value 'A' or 'B' |
| UFP rating | This field contains an ISBSG rating code of A, B, C or D applied to the Functional Size (Unadjusted Function Point) | Filtered record based on value 'A' or 'B' |
| Total defects delivered | Total defects delivered in the first month of use of the software | Not null values |

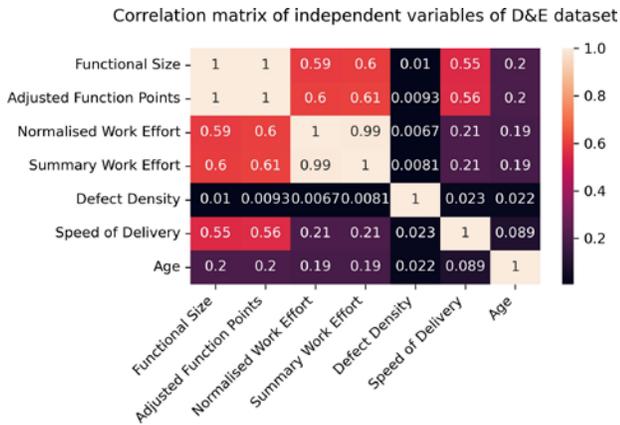

Fig. 2. Correlation Matrix for non-categorical features.

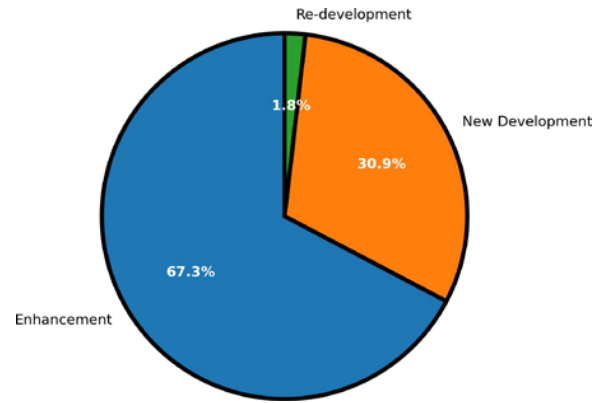

Fig. 4. Distribution of development types.

records with 10 independent variables, as described in Fig. 1, and one dependent or target variable namely "total defects delivered" field.

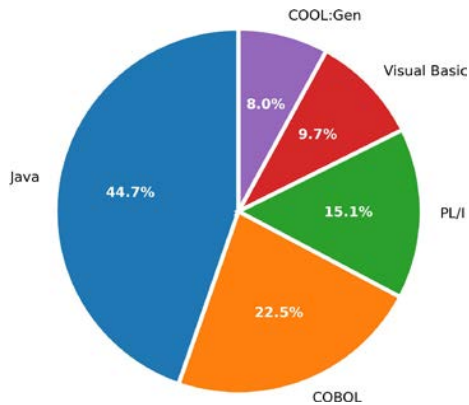

Fig. 3. Distribution of programming languages.

Next, the remaining records were further analyzed. The distribution of programming languages revealed that out of top 5 prominent programming languages, JAVA covered 44% of the projects as shown in Fig. 3. Additionally, the majority of the projects had a development type of enhancements among 67.3% of the total number of projects, followed by new development and redevelopment as shown in Fig. 4.

As a preprocessing step, all categorical values were converted to numerical values using the label encoder from the Scikit-learn library of Python [31]. Next, the dataset was split into 70% for training and 30% for training and testing.

We applied standardized scaling to the predictors. StandardScaler removes the mean and scales each feature to unit variance. This operation is performed feature-wise [32].

### C. Applied machine learning algorithms

In 2020, citeb33 utilized ensemble tree-based machine learning algorithms for SDP and obtained acceptable results on classification problems. In this study, we have applied various tree-based regression-based machine learning techniques as tree-based algorithms are popular for regression problems. Several of these algorithms have already been applied in classification [33], and a few have been used for regression problems in SDP models in the existing literature. The selected existing tree-based machine learning algorithms utilized for our evaluation have been listed below: **Random Forest Regression:** Random Forest [34] is a combination of tree predictors where each tree depends on the values of a random vector sampled independently with the same distribution for all trees in the forest. **AdaBoost Regression:** AdaBoost Regression [35] is a meta-estimator that begins by fitting a regressor on the original dataset and then fits additional copies of the regressor on the same dataset. **Gradient Boosting Regression:** GBRT [36] is a flexible non-parametric statistical learning technique for regression. **Extra Tree Regression:** Proposed by Geurts et al. [37] in 2006, Extremely Randomize Tree algorithm is a tree-based method that implements a meta estimator which fits a number of randomized decision trees on various sub-samples of the dataset and uses averaging to improve the predictive accuracy and control over-fitting. **XGBoost Regression:** XGBoost is an efficient implementation of gradient boosting [38].**Catboost Regressor** CatBoost is a ML algorithm that uses gradient boosting on decision trees. [33] Finally, **SHAP** [39], a game-theory-based approach to explain the output of the SDP models, was applied.

### D. Evaluation Criteria

For the evaluation of the SDP models, we applied several commonly used evaluation metrics for regression models in both training and testing datasets. Three of these metrics are Mean Absolute Error (MAE), Mean Squared Error (MSE) and Root Mean Squared Error (RMSE). These measures have been applied in various defect prediction studies [40], [41]. **MAE** is calculated as the sum of absolute errors divided by the sample size, representing the difference between predicted and actual value [41]. **MSE** represents the average of the squared difference of predicted and actual value. **RMSE** measures the standard deviation of the predicted errors which is the squared root of the MSE. To evaluate how well the developed SDP models explain the dataset, we also used state-of-the-art evaluation metrics $R^2$ and the adjusted $R^2$ values [42]. $R^2$ can be defined as the proportion of the total variation in the dependent variable that is explained by the independent variables. Adjusted $R^2$ is a modification of $R^2$ which adjusts for the number of explanatory terms. The difference between R-squared and adjusted R-squared value is that R-squared value assumes that all the independent variables considered affect the model, whereas the adjusted R squared value considers only those independent variables that influence the performance of SDP models.

## IV. RESULTS

In this section, we present the results of the empirical evaluation conducted to address the research questions. This study was performed on a dataset containing 1254 projects, which included projects developed in different programming languages. The analysis was carried out using Anaconda Navigator, and Python version 3.9.7. We used various Python libraries including scikit-learn ensemble and other relevant packages.

We verified the correlation between the defect density and total number of defects, as shown in Fig. 5. This figure demonstrates a strong positive relationship between defect density and the total number of defects. The correlation coefficient has a statistical value of 0.7863112869053916 with a p-value of $4.223627046436646e^{-264}$. This indicates that defect density has a highly significant relationship with the number of defects. Next, the correlation between functional size and total number of defects were verified as depicted in Fig. 6. These two variables also show a positive relationship. The Pearson correlation metrics [43] confirms a significant relationship between the size of the project and the number of defects. The Pearson correlation has a regression coefficient value of 0.29419230162036336, and p-value of $1.861125251801106e^{-26}$.

The results of the applied algorithms are shown in Fig. 7. The classifiers were evaluated on both training and testing data. For hyperparameter tuning, RandonSearchCV was utilized. Afterwards, a 5-fold cross- validation was applied for each of these algorithms. As expected, the training score was higher than the testing score in all models. Although GradientBoostingRegressor and XGBRegressor performed the best on the training dataset with R-squared, and adjusted R-squared values, as well as the lowest MAE and MSE, and

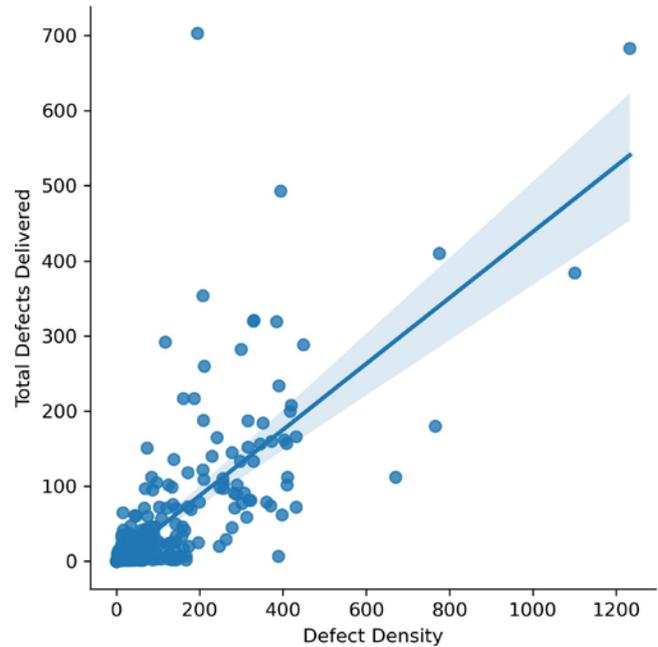

Fig. 5. Correlation between defect density and number of defects.

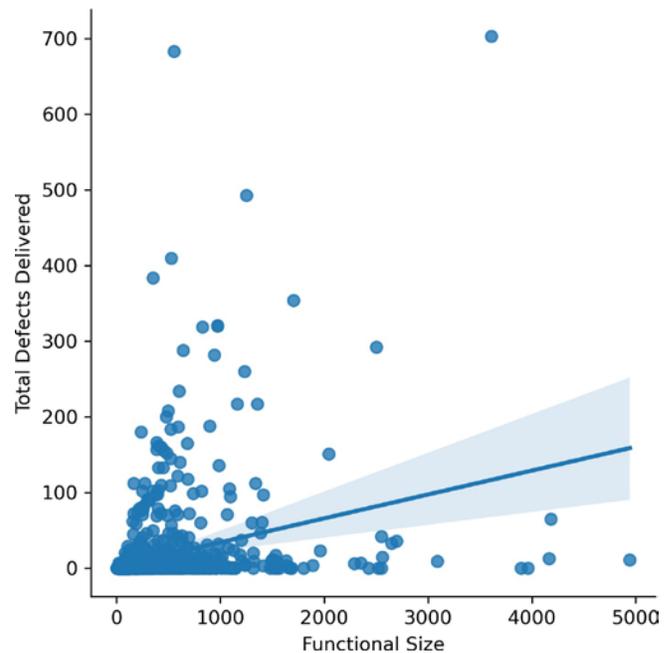

Fig. 6. Correlation between functional size and number of defects.

| ModelName | Cross validation score | Training R-squared value | Training adjusted R-squared value | Training MAE | Training MSE | Training RMSE | Testing R-squared value | Testing adjusted R-squared value | Testing MAE | Testing MSE | Testing RMSE |
|---|---|---|---|---|---|---|---|---|---|---|---|
| AdaBoostRegressor | 0.79254 | 0.95017 | 0.94959 | 7.18556 | 126.804 | 11.26073 | 0.802270637 | 0.796868195 | 9.01022 | 464.911 | 21.5618 |
| *CatBoostRegressor | **0.92071** | 0.99992 | 0.99992 | 0.3148 | 0.212 | 0.460753 | 0.862033265 | 0.858263682 | 3.79063 | 324.3942 | 18.01095 |
| RandomForestRegressor | **0.85819** | 0.97065 | 0.97031 | 1.2765 | 74.690 | 8.64234 | 0.863646901 | 0.859921407 | 4.28531 | 320.6001 | 17.90531 |
| **ExtraTreeRegressor | **0.93055** | 0.99984 | 0.99984 | 0.32469 | 0.404 | 0.635726 | 0.893156518 | 0.890237297 | 3.73382 | 251.2157 | 15.84978 |
| XGBoostRegressor | 0.81262 | 1 | 1 | 0.00019 | 0.000 | 0.000383 | 0.838245308 | 0.83382578 | 4.36809 | 380.3256 | 19.50194 |
| GradientBoostRegressor | 0.78644 | 1 | 1 | 0.00095 | 0.000 | 0.003904 | 0.846245282 | 0.842044334 | 4.34469 | 361.5157 | 19.01357 |

Fig. 7. Results from applied tree-based machine learning classifiers.

| | Feature_names | AdaBoostRegressor | CatBoostRegressor | RandomForestRegressor | ExtraTreeRegressor | XGBRegressor | GradientBoostingRegressor |
|---|---|---|---|---|---|---|---|
| 0 | Industry Sector | 0.017156 | 3.097268 | 0.016470 | 0.012470 | 0.005110 | 0.043050 |
| 1 | Development Type | 0.001878 | 3.149867 | 0.003090 | 0.003790 | 0.073210 | 0.000200 |
| 2 | Primary Programming Language | 0.003136 | 1.628474 | 0.002240 | 0.003800 | 0.019070 | 0.000340 |
| 3 | Count Approach | 0.029567 | 2.741773 | 0.000980 | 0.014630 | 0.091060 | 0.001070 |
| 4 | Functional Size | 0.290964 | 20.110008 | 0.225940 | 0.215000 | 0.178740 | 0.292350 |
| 5 | Relative Size | 0.011812 | 1.292116 | 0.023770 | 0.044310 | 0.086270 | 0.034980 |
| 6 | Normalised Work Effort | 0.085514 | 6.733606 | 0.063090 | 0.067000 | 0.027350 | 0.022380 |
| 7 | Defect Density | 0.541032 | 59.145665 | 0.656170 | 0.628970 | 0.491450 | 0.604210 |
| 8 | 1st Language | 0.012358 | 1.113497 | 0.002320 | 0.004020 | 0.000110 | 0.000430 |
| 9 | Age | 0.006584 | 0.987727 | 0.005940 | 0.006010 | 0.027630 | 0.001000 |

Fig. 8. Feature importance from each of the classifiers.

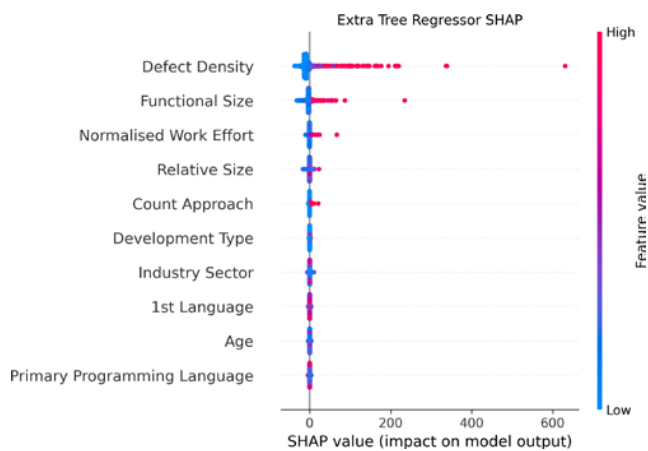

Fig. 9. ExtraTreesRegressor model explainability using SHAP.

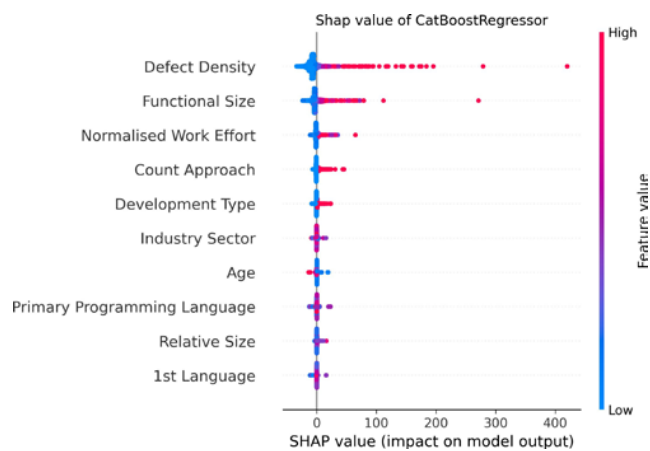

Fig. 10. CatBoostRegressor model explainability using SHAP.

RMSE, they did not do equally well on cross validation and testing data. This suggests that these models might have been overfitted during training. The ExtraTreeRegressor classifier shows $R^2$ score of 93% during 5-fold cross-validation. The testing dataset also exhibits a relatively high $R^2$ and adjusted $R^2$ score of 89% accompanied by the lowest testing MAE score of 3.7, MSE of 251 and RMSE of 15.8 among the applied algorithms. Since this classifier demonstrates the lowest error, as well as the highest $R^2$ and adjusted $R^2$ values during cross-validation, and testing, the ExtraTreeRegressor model is considered the most efficient model among the other models utilized in this study. The next efficient model is the CatBoostRegressor model with test dataset $R^2$ and adjusted $R^2$ values of 86%, and 85%, a cross-validation $R^2$ score of 92%, and lowes MAE value among the chosen models for this study. The RandomForest model also performed well in cross-validation and testing datasets. However, the AdaboostRegressor model performed relatively poor compared to the other selected classifiers, with a cross-validation $R^2$ score of approximately 79%, and slightly higher MAE, MSE and RMSE values than the other classifiers. Best on this analysis, the best performing model was the ExtraTreeRegressor followed by the CatBoostRegressor, and RandomForestRegressor.

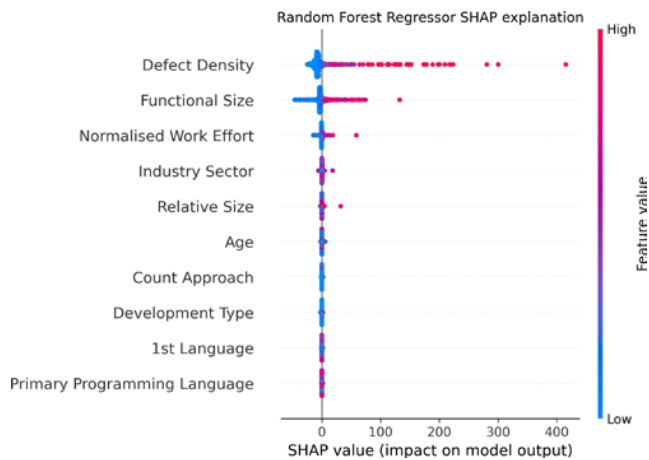

Fig. 11. RandomForestRegressor model explainability using SHAP.

As the next step, we recorded the feature importance for each of the applied algorithms as illustrated in Fig. 8. The top 5 important features are represented by the green color for each algorithms, while the minimum 3 performers are shown in red. Defect density emerged as the strongest feature in all 6 classifiers, followed by functional size. For instance, in the ExtraTreeRegressor model, the defect density feature had a coefficient of 63% while the coefficient for the 1st language was only 0.0047. This indicates that, for this SDP model, the 1st language feature did not provide a significant role ine the prediction. This field scored the lowest among all models, except for AdaBoostRegressor, but it was not among the top 5 predictors for AdaBoostRegressor either. Functional size was the second strongest predictor among all these classifiers. It seems that programming language for the software project did not have a significant impact in any of these predictions. On the other hand, "normalized work effort" contributed in all these models, although this feature had less importance compared to defect density and functional size.

To verify if these features can be explained using model agnostic approach, we applied SHAP on the top 3 models. The results are shown in Fig. 9, Fig. 10, and in Fig. 11 for the ExtraTreeRegressor, CatBoostRegressor and RandomForestRegressor models respectively. All these models aligned with the top three predictors in the same order which are defect density, functional size, and normalized work effort. However, for the 4th predictor, each model selected a different predictor. ExtraTreeREgressor identified Relative Size as the next important predictor, Catboost Regressor highlighted the counting approach, and RandomForestRegressor included Industry Sector. It appears that the programming language did not significantly contribute to the prediction of this SDP model.

## V. ANALYSIS AND DISCUSSION

This study employed regression models including Extra Randomized Tree, CatBoost, RandomForest, XGBoost, AdaBoost, and GradientBoosting algorithms to predict the number of defects. The findings revealed that ExtaTree, CatBoost, and RandomForest demonstrated better performance in this regard. The reliability on these models was achieved as they exhibited comparatively lower MAE, MSE, RMSE values in training and testing dataset, along with comparatively higher $R^2$ and adjusted $R^2$ scores. The application of SHAP models provided reliable explanations of the features, and the top three models show consistency in their top three predictors. This contributes towards explainable SDP models. While SHAP has been used in recent SDP studies [44], to the best of our knowledge, direct application of SHAP models in cross-company project datasets for model explanations has not been widely explored.

## VI. THREATS TO VALIDITY

This research was conducted on the ISBSG dataset. However, many of the records had to be removed due to missing values. Although the analysis was performed on a significant number of records, it is worth noting that adding the missing records could potentially alter the findings if information were readily available.

## VII. CONCLUSION AND FUTURE WORK

This study utilized six supervised tree-based ML algorithms for developing SDP models using cross-company project information from the ISBSG dataset. By employing regression models, several findings were derived from the selected attributes based on software size, work effort, defect density, development type, organization type, and primary programming language etc. This study reaffirmed the promising value of regression ML in SDP predictions. Furthermore, the feature importance of the selected attributes was observed, and correlation between the predictors and the number of defects were identified. Finally, the SDP models were explainable.

Future studies could adopt a more targeted approach by categorizing the dataset based on programming language, development type, and other factors to obtain more specific outcome in addition to this generic model presented here. Additionally, the selected attributes can be applied to other open-source projects to increase the reliability of the explained predictors after being validated by explainable machine learning models.

ACKNOWLEDGMENT

We would like to thank ISBSG for providing us with the data subscription. Also, we would like to acknowledge the support by Mrs. Mary Pierce, Dean of Faculty of Business,

## REFERENCES

[1] T. Elfriede and B. Garrett, Implementing Automated Software Testing How to Save Time and Lower Costs While Raising Quality, Addison-Wesley, 2009.

[2] M. Shepperd, Q. Song, Z. Sun, and C. Mair, "Data quality: Some comments on the NASA software defect datasets", in IEEE Trans. Softw. Eng., vol. 39, no. 9, Sep. 2013, pp. 1208–1215.

[3] S. Aleem, L. F. Capretz, and F. Ahmed, "Benchmarking machine learning techniques for software defect detection", in Int. J. Softw. Eng. Appl., vol. 6, May. 2015, pp. 11-23.

[4] A.O. Balogun, S. Basri, L.F. Capretz, S. Mahamad, A.A. Imam, M.A. Almomani, V.E. Adeyemo, A.K. Alazzawi, A.O. Bajeh, G. Kumar, "Software defect prediction using wrapper feature selection based on dynamic re-ranking strategy", in J. Symmetry, Vol.13, Issue 11, 2021, pp. 1-23, Paper 2166.

[5] J. Bai, J. Jia, L.F. Capretz, "A three-stage transfer learning framework for multi-source cross-project software defect prediction", in J. Inf. Softw. Technol., Vol. 150, 16 pages. 2022.

[6] A.O. Balogun, S. Basri, S. Mahamad, L.F. Capretz, A.A. Imam, M.A. Almomani, V.E. Adeyemo, G. Kumar, "A novel rank aggregation-based hybrid multifilter wrapper selection method in software defect prediction", in J. Computational Intelligence and Neuroscience, Vol. 2021, 2021, pp. 1-19, Paper 5069016.

[7] T. Menzies, J. DiStefano, A. Orrego, and R. Chapman, "Assessing predictors of software defects", in Proc. Workshop Predictive Softw. Models, 2004.

[8] C. Shan, B. Chen, C. Hu, J. Xue, and N. Li, "Software defect prediction model based on LLE and SVM", in Proc. Commun. Secur. Conf. (CSC '14), 2014, pp. 1–5.

[9] A. Rahim, Z. Hayat, M. Abbas, A. Rahim and M. A. Rahim, "Software Defect Prediction with Naïve Bayes Classifier", in 2021 Intl. Bhurban conf. on Applied Sciences and Technologies (IBCAST), Islamabad, Pakistan, 2021, pp. 293-297, doi: 10.1109/IBCAST51254.2021.9393250.

[10] M. A. Khan, N. S. Elmitwally, S. Abbas, S. Aftab, M. Ahmad, M. Fayaz, and F. Khan, "Software defect prediction using Artificial Neural Networks: A systematic literature review", in J. Sci. Program., vol. 2022, May. 2022, pp. 1–10,. "

[11] P. Deep Singh and A. Chug, "Software defect prediction analysis using machine learning algorithms," in Proc. 7th Int. Conf. Cloud Comput., in J. Data Sci. Eng. (Confluence), Jan. 2017, pp. 775–781." "

[12] Y. N. Soe, P. I. Santosa, and R. Hartanto, "Software defect prediction using random forest algorithm," in Proc. 12th South East Asian Tech. Univ. Consortium (SEATUC), Mar. 2018, pp. 1–5." "

[13] Z. Xu, J. Liu, X. Luo, Z. Yang, Y. Zhang, P. Yuan, Y. Tang, and T. Zhang, "Software defect prediction based on kernel PCA and weighted extreme learning machine", in Inf. Softw. Technol., vol. 106, Feb. 2019, pp. 182–200."

[14] W. Wu, C. Feng, H. Ren, X. Han, and X. Tong, "Research on software defect prediction system based on deep learning", in Fifth Intl. Conf. Mechatronics and Comput. Technol. Eng. (MCTE 2022), Chongqing, China, 2022.

[15] A. Ho, N. Nhat Hai, and B. Thi-Mai-Anh, "Combining deep learning and kernel PCA for software defect prediction", in Proc. 11th Intl. Symp. Inf. and Commun. Tech., 2022, pp. 360-367.

[16] N. Massoud and P. Kumar Jain, "Software defect prediction using dimensionality reduction and deep learning", in Third Intl. conf. on Intelligent Commun. Technol. and Virtual Mobile Networks (ICICV), 2021, pp. 884-893.

[17] H. Issam, M. Alshayeb, and L. Ghouti, "Software defect prediction using ensemble learning on selected features", in Info. and Softw. Technologies, vol. 58, 2015, pp. 388–402.

[18] M. W. Thant and N. T. T. Aung, "Software Defect Prediction using Hybrid Approach", in 2019 Intl. conf. on Advanced Information Technologies (ICAIT), Yangon, Myanmar, 2019.

[19] M. Ali and S. Awais Mian, "Improving Recall of software defect prediction models using association mining", Knowledge-Based Systems, vol. 90, 2015, pp. 1–13, .

[20] G. K. Rajbahadur, S. Wang, Y. Kamei, and A. E. Hassan, "The impact of using regression models to build defect classifiers", arXiv [cs.SE], Feb. 12, 2022.

[21] E. A. Felix and S. P. Lee, "Integrated approach to software defect prediction",in IEEE Access, vol. 5, 2017, pp. 21524–21547.

[22] B. Gezici and A. K. Tarhan, "Explainable AI for Software Defect Prediction with Gradient Boosting Classifier," in 7th Intl. Conf. Comput. Sci. and Eng. (UBMK), Diyarbakir, Turkey, 2022, pp. 1-6, doi: 10.1109/UBMK55850.2022.9919490.

[23] M. Almakadmeh and A. Abran, "The ISBSG software project repository An analysis from Six Sigma measurement perspective for software defect estimation", in J. Soft. Eng. and Appl., vol. 10, no. 8, 2017, pp. 693–720.

[24] S. Fadi and W. Al-Manai, "Toward An Empirical Study to investigate the size-defect Relationship Using ISBSG Repository", in Proc. Intl. Conf. on Intelligent Information Processing, Security and Advanced Communication, pp. 1-5, 2015

[25] Z. He, F. Shu, Y. Yang, M. Li, and Q. Wang, "An investigation on the feasibility of cross-project defect prediction", in J. Automated Soft. Eng., vol. 19, no. 2, Jun. 2012, pp. 167–199.

[26] Y. Shao, J. Zhao, X. Wang, W. Wu, and J. Fang, "Research on Cross-Company Defect Prediction Method to Improve Software Security", in J. Secur. and Commun. Netw., vol. 2021, 2021, pp. 1–19.

[27] J. Shin, R. Aleithan, J. Nam, J. Wang, and S. Wang, "Explainable software defect prediction: Are we there yet?", arXiv preprint arXiv:2111. 10901, 2021. "

[28] The ISBSG Development & Enhancement project data , ISBSG, R 21, Sep. 2021, http://www.isbsg.org

[29] ISBSG D & E - Release Demographics Sept 2021 R2, ISBSG, 2021.

[30] Guidelines for use of the ISBSG data, ISBSG, 2021

[31] Field Descriptions ISBSG D&E Repository, ISBSG, Release 2021.

[32] Pedregosa et al., "Scikit-learn: Machine learning in Python", in J. Machine Learning Res., vol. 12, 2011, pp. 2825–2830.

[33] H. Aljamaan and A. Alazba, "Software Defect Prediction Using Tree-Based Ensembles", in Proc. 16th ACM Intl. Conf. on Predictive Models and Data Analytics Soft. Eng, Virtual, USA, 2020, pp. 1–10.

[34] L. Breiman, "Random forests", in Machine learning, vol. 45, 2001 pp. 5–32.

[35] D. P. Bibitemb Solomatine and D. L. Shrestha, "AdaBoost. RT: a boosting algorithm for regression problems", in Proc. IEEE Intl. Joint Conf. on Neural Netw., vol. 2, IEEE, 2004, vol. 2, pp. 1163–1168.. "

[36] I. H. Laradji, M. Alshayeb, and L. Ghouti, "Software defect prediction using ensemble learning on selected features", in Inf. Technol., vol. 58, pp. 388–402, Feb. 2015, doi: 10.1016/j.infsof.2014.07.005."

[37] R. Mareé, L. Wehenkel, and P. Geurts, "Extremely randomized trees and random subwindows for image classification, annotation, and retrieval", in Decision Forests for Computer Vision and Medical Image Analysis, London: Springer London, 2013, pp. 125–141.

[38] T. Chen and C. Guestrin, "Xgboost: A scalable tree boosting system", in Proceedings 22nd acm sigkdd Intl. conf. on Knowl. discovery and data mining, 2016, pp. 785–794.

[39] G. Esteves, E. Figueiredo, A. Veloso, M. Viggiato, and N. Ziviani, "Understanding machine learning software defect predictions", in J. Autom. Softw. Eng., vol. 27, no. 3–4, pp. 369–392, Dec. 2020.

[40] F. Yang, Y. Huang, H. Xu, P. Xiao, and W. Zheng, "Fine-Grained Software Defect Prediction Based on the Method-Call Sequence", Computational Intelligence and Neuroscience, vol. 2022, 2022.

[41] A. Botchkarev, "A new typology design of performance metrics to measure errors in machine learning regression algorithms", in Interdiscip. J. Inf. Knowl. Manag., vol. 14, 2019, pp. 045–076,.

[42] X. Tan, X. Peng, S. Pan, and W. Zhao, "Assessing software quality by program clustering and defect prediction", in 2011 18th Working conf. on Reverse Engineering, Limerick, Ireland, 2011.

[43] F. Coelho, A. P. Braga, and M. Verleysen, "Multi-Objective Semi-Supervised Feature Selection and Model Selection Based on Pearson's Correlation Coefficient," Lecture Notes in Computer Science, vol. 6419, no. 11, pp. 509-516, 2010.

[44] B. Gezici and A. K. Tarhan, "Explainable AI for Software Defect Prediction with Gradient Boosting Classifier", in 2022 7th Intl. conf. on Computer Science and Engineering (UBMK), 2022, pp. 1–6.